\documentclass{ICASP13Paper}
\usepackage{amssymb}
\usepackage{amsmath}
\addauthor{Saeed Nozhati}{Graduate Student, Dept. of Civil and Environmental Engineering, Colorado State University, Fort Collins, USA}
\addauthor{Yugandhar Sarkale}{Graduate Student, Dept. of Electrical and Computer Engineering, Colorado State University, Fort Collins, USA}
\addauthor{Bruce R. Ellingwood}{Professor, Dept. of Civil and Environmental Engineering, Colorado State University, Fort Collins, USA}
\addauthor{Edwin K.P. Chong}{Professor, Dept. of Electrical and Computer Engineering, Colorado State University, Fort Collins, USA}
\addauthor{Hussam Mahmoud}{Associate Professor, Dept. of Civil and Environmental Engineering, Colorado State University, Fort Collins, US}

\title{An approximate dynamic programming approach to food security of communities following hazards}

\referencefile{bib}

\shortabstract{Food security can be threatened by extreme natural hazard events for households of all social classes within a community. To address food security issues following a natural disaster, the recovery of several elements of the built environment within a community, including its building portfolio, must be considered.  Building portfolio restoration is one of the most challenging elements of recovery owing to the complexity and dimensionality of the problem. This study introduces a stochastic scheduling algorithm for the identification of optimal building portfolio recovery strategies. The proposed approach provides a computationally tractable formulation to manage multi-state, large-scale infrastructure systems.  A testbed community modeled after Gilroy, California, is used to illustrate how the proposed approach can be implemented efficiently and accurately to find the near-optimal decisions related to building recovery following a severe earthquake.}

\begin{document}
     One of the principal objectives of the United Nations (UN) Sustainable Development Goals is achieving food security. The Food and Agriculture Organization (FAO) describes food security as: "a situation that exists when all people, at all times, have physical, social and economic access to sufficient, safe and nutritious food that meets their dietary needs and food preferences for an active and healthy life" (\cite{FAO}). Securing an adequate food supply to all community inhabitants requires a food distribution system that is resilient to natural and man-made hazards.  The growth of population in hazard-prone regions  and climate change pose numerous challenges to achieving a resilient food system around the world. The resiliency concept applied to food distribution systems can be evaluated with respect to two different time-frames, namely in "normal" times (i.e., prior to disasters) and in the aftermath of hazards. Several studies have investigated different approaches to enhance the resilience of agri-food systems (\cite{seekell}). These studies have focused on resilience in terms of biophysical capacity to increase food production, diversity of modern domestic food production, and the role played by social status and income in the impact of food deficits.  To mitigate food security issues, the United States Department of Agriculture (USDA) Food and Nutrition Service (FNS) supplies 15 domestic food and nutrition assistance programs. The three largest are the Supplemental Nutrition Assistance Program (SNAP - formerly the Food Stamp Program), the National School Lunch Program, and the Special Supplemental Nutrition Program for Women, Infants, and Children (WIC) (\cite{Oliv}). However, household food security following extreme natural hazard events is also contingent on interdependent critical infrastructure systems, such as transportation, energy, water, household units, and retailer availability.
     \par
     This study focuses on the connection between failures in food distribution and food retail infrastructure and disruption in civil infrastructure and structures. Household food security issues are considerably worsened following natural disasters. For example, Hurricanes Rita, Wilma, and Katrina, which occurred in 2005, caused disaster-related food programs to serve 2.4 million households and distributed \$928 million in benefits to households  (\cite{Food}). Three dimensions of food security - accessibility, availability, and affordability - are particularly relevant for the nexus between infrastructure and household food security.  Affordability captures the ability of households to buy food from food retailers, and is a function of household income, assets, credit, and perhaps even participation in food assistance programs. Accessibility is concerned with the households’ physical access to food retail outlets.  Because at least one functional route must be available between a household unit and a functioning food retailer, transportation networks are a major factor in accessibility. Availability is concerned with the functionality of the food distribution infrastructure, beginning with wholesalers, extending to retailers, and ultimately ending with the household as the primary consumer. The functionality of food retailers and household units depends not only on the functionality of their facilities but also the availability of electricity and water. Therefore, the electrical power network (EPN), water network (WN), and the buildings housing retailers and household units must be considered simultaneously to address availability.
\par
As is evident from the preceding discussion, food security relies on a complex supply-chain system. If such a system is disrupted, community resilience and the food security will be threatened (\cite{paci}). In this paper, we focus only on household unit structures, which forms the largest entity in community restoration. In this paper, we focus on household unit buildings, which usually form the largest element of the built environment in community restoration. A literature review (\cite{Wang}) shows that the recovery of building portfolios has been studied far less than the recovery of other infrastructure systems.  Building portfolio restoration is an essential element of availability and plays a major role towards addressing food security issues.  Effective emergency logistics demand a comprehensive decision-making framework that addresses and supports policymakers' preferences by providing efficient recovery plans. In this study, we employ Markov decision processes (MDPs) along with an approximate dynamic programming (ADP) technique to provide a practical framework for representation and solution of stochastic large-scale decision-making problems. The scale and complexity of building portfolio restoration is captured by the proposed simulation-based representation and solution of the MDP. The near-optimal solutions are illustrated for the building portfolio of a testbed community modeled after Gilroy, California, United States.

\section{TESTBED CASE STUDY}
\begin{figure}[h!]
	\centering
	\includegraphics[scale=0.6]{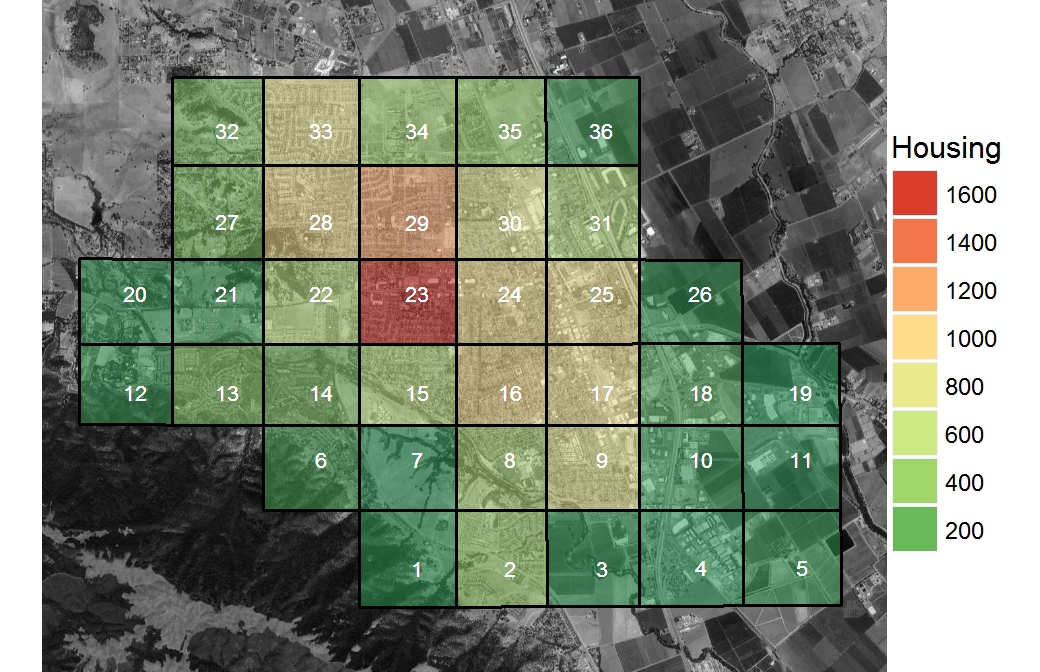}
	\caption{Housing units over the defined grids. }
	\label{fig1}
\end{figure}

As an illustration, this study considers the building portfolio of Gilroy, California, USA. The City of Gilroy is a moderately  sized growing city in southern Santa Clara County, California, with a population of 48,821 at the time of the 2010 census.  The study area is divided into 36 rectangular regions organized as a grid to define the properties of the community with an area of 42 $  km^{2} $ and a population of 47,905. Household units are growing at a faster pace in Gilroy than in Santa Clara County and the State of California  (\cite{Harnish}). The average number of people per household in Gilroy in 2010 was 3.4, greater than the state and county average.  Approximately 95\% of Gilroy's housing units are occupied. A heat map of household units in the grid is shown in Figure~\ref{fig1}. Age distribution of Gilroy is tabulated in Table~\ref{tabl1}.

\begin{table}[h!]
	\caption{Age distribution of Gilroy (\cite{Harnish}).}
	\label{tabl1}
	\centering
	\begin{tabular}{|c|c|}
		\hline
		Age Group  & Percent \\
		\hline
		Children (0-17 years) & 30.60 \\
		\hline
		Adults (18-64 years)  & 61 \\
		\hline
		Senior Citizen (65+ years)  & 8.40 \\
		\hline
	\end{tabular}
\end{table}
\par

\section{SEISMIC HAZARD AND DAMAGE ASSESSMENT}
     The seismic hazard is a dominant hazard of California. Hence, we consider a seismic event of moment magnitude $ M_{w} = 6.9 $ that occurs at one of the closest points on the San Andreas Fault  to downtown Gilroy with an epicentral distance of approximately 12 km. We used the \cite{abrahamson} ground motion prediction equation (GMPE) to evaluate the Intensity Measures (IM) and associated uncertainties, including the intra-event (within event) and inter-event (event-to-event) uncertainties, at the sites of each of the 14,702 buildings in Gilroy.  We assessed the damage to household units and food retailers with the seismic fragility curves presented in HAZUS-MH (\cite{fema}). We considered repair vehicles, crews, and tools as available \textit{resource units} (RUs) to restore the buildings following the hazard. One RU is required to repair each damaged building (\cite{masoomi}). We adopted the synthesized restoration time from HAZUS-MH.

\section{MARKOV DECISION PROCESS FRAMEWORK}
We provide a brief description of MDPs; additional details of MDPs are available elsewhere (\cite{puterman}). A MDP is defined by the five-tuple $(X,A,P,R,\gamma)$, where $X$ denotes the state space, $A$ denotes the action space, $P(y|x,a)$ is the probability of transitioning from state $x\in X$ to state $y\in Y$ when action $a$ is taken, , $ R(x, a) $ is the reward obtained when action $ a $ is taken in state $x\in X$, and $ \gamma $ is the discount factor. A policy $ \pi: X\longrightarrow A $ is a mapping from states to actions, and $\varPi $ be the set of policies ($ \pi $). The objective is then to find the optimal policy, denoted by $ \pi^{*} $, that maximizes the total reward (or minimizes the total cost) over the time horizon, i.e.,

\begin{equation}\label{eq1}
\pi^*:=\arg\sup_{\pi \in \varPi} V^{\pi}(x),
\end{equation}
where
\begin{equation}\label{eq2}
V^\pi(x):=E\left\lbrack\sum_{t=0}^{\infty}\gamma^{\,t}R(x_t,\pi(x_t))|x_{0}=x\right\rbrack,
\end{equation}

$ V^{\pi}(x) $ is called the value function for a fixed policy $ \pi $, and $ 0<\gamma<1 $ is the discount factor (\cite{puterman}).
The \textit{optimal value function} for a given state $x\in X$ is connoted as $ V^{\pi^{*}}(x): X\longrightarrow \mathbb{R} $ given by
\begin{equation}\label{eq3}
V^{\pi^{*}}(x):=\sup_{\pi \in \varPi} V^{\pi}(x).
\end{equation}

Bellman’s optimality principle (\cite{bertsekas}) is useful for defining \textit{Q}-value function. \textit{Q}-value function plays a pivotal role in the description of the rollout algorithm. Bellman’s optimality principle states that  $ V^{\pi^{*}}(x)$ satisfies
\begin{equation}\label{eq4}
V^{\pi^{*}}(x):=\sup_{a \in A(x)}\left\lbrack R(x,a)+\gamma \sum_{y \in X} P(y|x,a)V^{\pi^{*}}(y) \right\rbrack,
\end{equation}
The $ Q $-value function associated with the optimal policy $ \pi^{*} $ is defined as
\begin{equation}\label{eq5}
Q^{\pi^{*}}(x,a):= R(x,a)+\gamma \sum_{y \in X} P(y|x,a)V^{\pi^{*}}(y),
\end{equation}
which is the inner-term on the R.H.S. in Eq.~(\ref{eq4}).
\par
Theoretically, $ \pi^{*} $ can be computed with linear programming or dynamic programming (DP). However, exact methods are not feasible for real-world problems that have large state and action spaces, like the community-level optimization problem considered herein, owing to the \textit{curse of dimensionality}; thus, an approximation technique is essential to obtain the solution. In the realm of approximate dynamic programming (ADP) techniques, a model-based, direct simulation approach for \textit{policy evaluation} is used (\cite{ieee}). This approach is called “rollout.” Briefly, an estimate $\hat{Q}^{\pi}(x,a)$ of the \textit{Q}-value function is calculated by Monte Carlo simulations (MSC) in the rollout algorithm as follows: we first  simulate $  N_{MC} $ number of trajectories, where each trajectory is generated using the policy $ \pi $ (called the \textit{base policy}), has length $ K $, and starts from the pair $(x, a)$; then, $\hat{Q}^{\pi}(x,a)$ is the average of the sample functions along these trajectories:
\begin{equation}\label{eq6}
\resizebox{0.5\textwidth}{!}{$\hat{Q}^{\pi}(x,a)=\dfrac{1}{N_{MC}} \sum_{i_{0}=1}^{{N_{MC}}}\left\lbrack R(x,a,x_{i_{0},1})+\sum_{k=1}^{{K}}\gamma^{k}R(x_{i_{0},k},\pi(x_{i_{0},k},x_{x_{i_{0},k+1}}))\right\rbrack.$}
\end{equation}
For each trajectory $ i_{0} $, we fix the first state-action pair to $ (x, a) $; the next state $ x_{i_{0},1} $ is calculated when the current action $ a $ in state $ x $ is completed. Thereafter, we choose actions using the base policy. A more complete description of the rollout algorithm can be found in (\cite{bertsekas,nozhati2019}).
\section{BUILDING PORTFOLIO RECOVERY}
Each household unit and retailer building remains undamaged or exhibits one of the damage states (i.e., Minor, Moderate, Major, and Collapse) based on the level of intensity measure and the seismic fragility curves. There is a limited number of RUs (defined earlier) available to the decision maker for the repair of the buildings in the community. In this study, we also limit the number of RUs for each urban grid so that the number of available RUs for each grid $RU_g$ is 20 percent of the number of damaged buildings in each region of the grid. Therefore, the number of RUs varies over the community in proportion to the density of the damaged buildings.
\par
Let $ x_{t} $ be the state of the damaged structures of the building portfolio at time $ t $; $ x_{t} $ is a vector, where each element represents the damage state of each building in the portfolio based on the level of intensity measure and the seismic fragility curves. Let $ a_{t}^{g} $ denote the repair action to be carried out on the damaged structures in the $ g^{th} $ region of the grid at time $ t $; each element of  $a_{t}^{g}$ is either zero or a one, where zero means do not repair and one means carry out repair. Note that the sum of elements of $a_{t}^{g}$ is equal to $RU_g$. The repair action for the entire community at time $ t $, $ a_{t} $, is the stack of the repair action $ a_{t}^{g} $. The assignment of RUs to damaged locations is $ non-preemptive $ in the sense that the decision maker cannot preempt the assigned RUs from completing their work and reassign them to different locations at every decision epoch $ t $. This type of scheduling is more suitable when the decision maker deals with non-central stakeholders and private owners, which is the case for a typical building portfolio. We wish to plan decisions optimally so that a maximum number of inhabitants have safe household unit structures per unit of time (day in our case). Therefore, the reward function embeds two objectives as follows:
\begin{equation}\label{eq7}
R(x_t,a_t,x_{t+1})=\frac{r}{t_{rep}},
\end{equation}
where $ r $ is the number of people benefited from household units after the completion of $ a_{t} $, and $ t_{rep} $ is the total repair time to reach $ x_{t+1} $ from any initial state $ x_{0} $. Note that the reward function is stochastic because the outcome of the repair action is stochastic. In this study, we set the discount factor to be 0.99, implying that the decision maker is “far-sighted” in the consideration of the future rewards.
\par
\begin{figure}[h!]
	\centering
	\includegraphics[width=.5\textwidth]{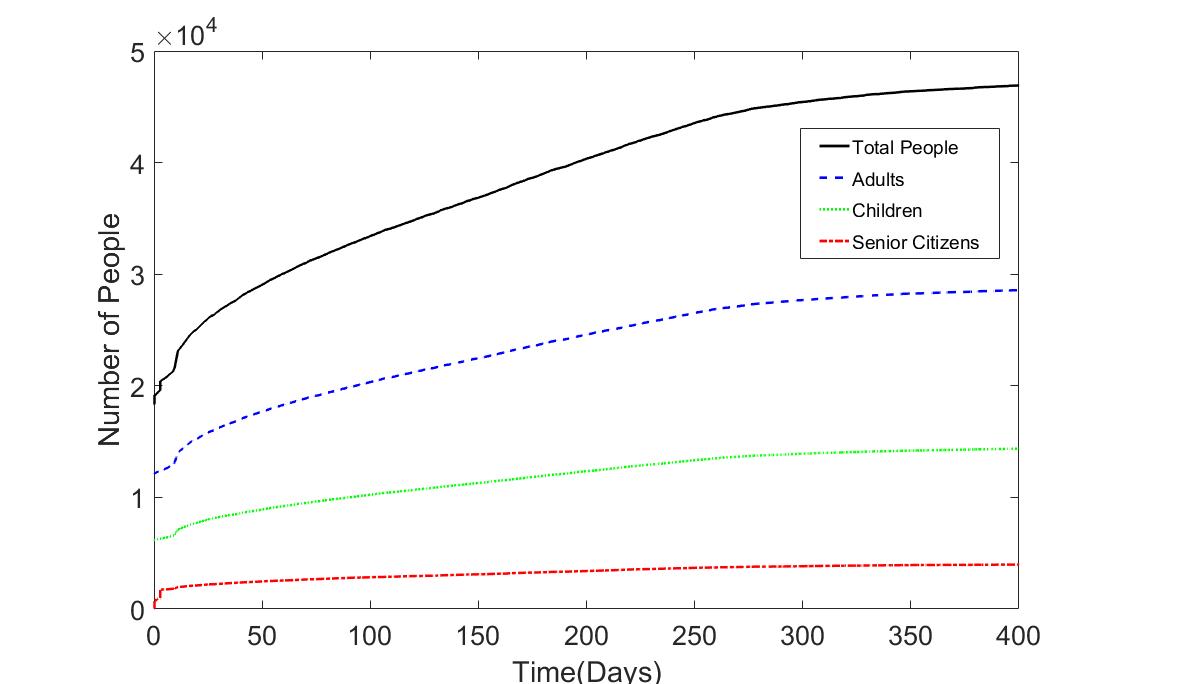}
	\caption{Different numbers of people based on age with inhabitable structures.}
	\label{fig2}
\end{figure}

We simulated $  N_{MC} $ number of trajectories to reach a low (0.1 in this study) dispersion in Eq.~(\ref{eq6}). As Eq.~(\ref{eq6}) shows, we addressed the mean-based optimization that is suited to risk-neutral decision-makers. However, this approach can easily address different risk aversion behaviors. Figure~\ref{fig2} shows the total number of people with inhabitable structures (undamaged or repaired) over the community. We also computed the different numbers of children, adults, and senior citizens that have safe buildings over the recovery.  Different age groups have different levels of vulnerability to food insecurity; for example, children are a vulnerable group and must be paid more attention during the recovery process.

\begin{figure}[h!]
	\centering
	\includegraphics[width=.5\textwidth]{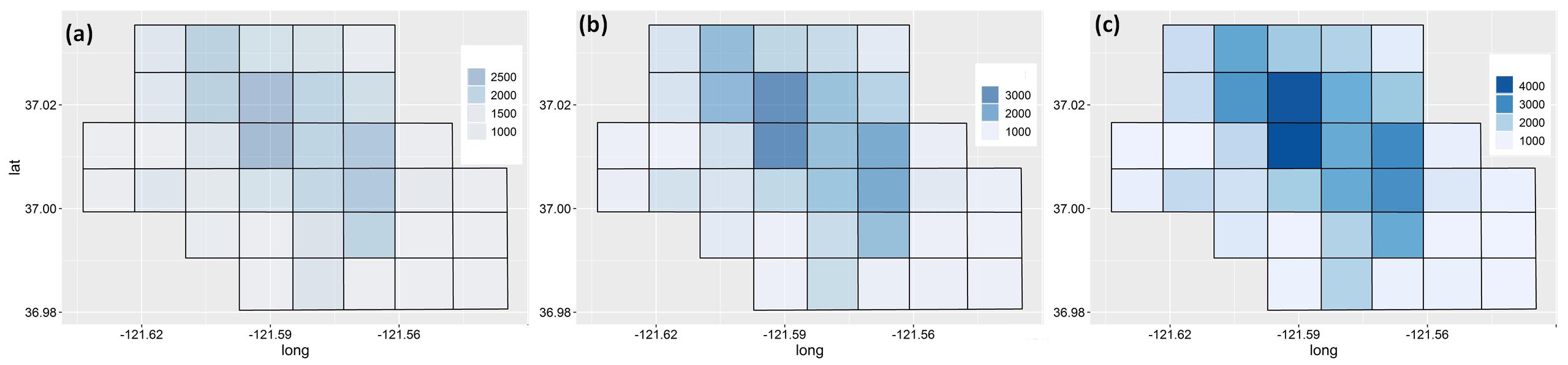}
	\caption{Number of people with inhabitable houses a) following the earthquake b) after 100 days c) after 600 days.}
	\label{fig3}
\end{figure}

Figure~\ref{fig3} depicts the spatio-temporal evolution of the community for people with inhabitable structurally-safe household units. This figure shows that for urban grids with a high density of damaged structures, complete recovery is prolonged despite availability of additional RUs. The spatio-temporal analysis of the community is informative for policy makers whereby they can identify the vulnerable areas of the community across time.

\section{CONCLUSION AND FUTURE WORK }
The building portfolio restoration is one of the most challenging ingredients to address food security issues in the aftermath of disasters. Our stochastic dynamic optimization approach, based on the method of rollout, successfully plans a near-optimal building portfolio recovery following a hazard. Our approach shows how to overcome the curse of dimensionality in optimizing large-scale building portfolio recovery post-diaster. For future work, we consider several aspects of a community from infrastructure systems to social systems along with their interdependencies. We will also explore how to fuse meta-heuristics to our solution to supervise the stochastic search that determines the most promising actions (\cite{nozhati2018}).
\section{REFERENCES}
{\small
	\bibliographystyle{ascelike}
	\bibliography{\bibfile}
}

\end{document}